\documentclass[twocolumn,showpacs,preprintnumbers,amsmath,amssymb]{revtex4}
\usepackage{amsmath,amsfonts,diagbox,latexsym,amssymb,graphics,epsfig,subfigure,color,makeidx}
\usepackage{xcolor}
\usepackage[utf8]{inputenc}
\usepackage{graphicx,hyperref}
\usepackage{bm}
\usepackage{color}
\usepackage{multirow}

\begin{document}
	
	\title{Dynamical formation of axionic hair around charged black hole}
	\author{Yu-Peng Zhang\footnote{zhangyupeng@lzu.edu.cn},
		Si-Jiang Yang\footnote{yangsj@lzu.edu.cn},
        Shao-Wen Wei\footnote{weishw@lzu.edu.cn},
        Wen-Di Guo\footnote{gwd@lzu.edu.cn},
		Yu-Xiao Liu\footnote{liuyx@lzu.edu.cn, corresponding author}
	}
	\affiliation{
        Key Laboratory of Quantum Theory and Applications of MoE, Lanzhou Center for Theoretical Physics, Key Laboratory of Theoretical Physics of Gansu Province, Lanzhou University, Lanzhou 730000, China\\
        Institute of Theoretical Physics \& Research Center of Gravitation, Lanzhou University, Lanzhou 730000, China
	}
	
	\begin{abstract}
		
	In this paper, we present a nonlinear numerical investigation on the dynamical scalarization process of a Reissner-Nordstr\"om black hole, incorporating an axionic scalar potential within the framework of the Einstein-Maxwell-dilaton theory. By scrutinizing the evolution of the irreducible mass of the black hole and the value of scalar field on the apparent horizon across various parameters of the axionic potential, we elucidate the correlations between the final states of scalarized charged black hole and the axionic potential. We observe that the inclusion of the axionic potential can either decrease or increase the irreducible mass of the final scalarized black hole, depending on the strength of the coupling between the dilation and the electric invariant $F_{\mu\nu}F^{\mu\nu}$. Regarding the value of the scalar field on the apparent horizon, we find that it decreases with the inclusion of the axionic potential. Our results contribute to an important understanding of the dynamical scalarization of black holes and the potential configurations of scalar hair with various self-interactions.
		
	\end{abstract}
	\maketitle
	\section{Introduction}

The recent observations of gravitational waves from the black hole mergers \cite{Abbott2016a,LIGOScientific:2018jsj,LIGOScientific:2020kqk,LIGOScientific:2021psn} and black hole shadows \cite{eth2019,eth2022} have opened up new avenues for testing and understanding the nature of black holes. To further investigate the properties of black holes based on astrophysical observations, it is essential to gain an accurate understanding of their characteristics. The uniqueness theorem has proved that the outcome of gravitational collapse of any type matter-energy is a Kerr-Newman black hole, and the corresponding properties of a stationary black hole in electro-vacuum general relativity are solely characterized by its mass $M$, charge $Q$, and angular momentum $J$ \cite{Chrusciel:2012jk}. However, the no-hair theorem of black hole would be violated when the non-Abelian hair or non-minimal coupling between the hair field and background spacetime is considered, such as the Yang-Mills hairy black holes \cite{Volkov:1989fi,Bizon:1990sr,Greene:1992fw}, the Skyrme hairy black holes \cite{Luckock:1986tr,Droz:1991cx}, and scalar dilaton hairy black holes \cite{Kanti:1995vq}.

As the simplest matter field, the scalar field is widely used to understand some unknown physical phenomena, especially for understanding the nature of dark matter \cite{Marsh:2015xka,Cardoso:2019rvt}, the inflation of cosmology \cite{Bezrukov:2007,Burgess2009,Burgess2010,Giudice2011,Lerner2010,Lyth1999}. The nonlinear simulation for the gravitational collapse of a scalar field also revealed the critical behavior in the formation of black holes \cite{Choptuik1993,Goncalves:1997qp,Healy:2013xia,Kidder2019}. Under the conformal transformation, the scalar field also provides the connects between general relativity and modified gravity theories \cite{Wagoner1970,Felice2010,Sotiriou2010}. Therefore, the investigation of the black hole solutions in static scalar-vacuum spacetimes constitutes a significant aspect as well \cite{Chase:1970omy,Bekenstein:1971hc,Bekenstein:1972ky,Bekenstein:1995un}.

In 1972, Chase firstly showed that there are no asymptotically flat static black hole solutions with minimally coupled massless scalar hair in general relativity \cite{Chase:1970omy}. Then Hawking and Bekenstein further proposed the no-hair theorem \cite{Hawking:1972qk,Bekenstein:1971hc,Bekenstein:1972ky,Bekenstein:1995un} and proved that minimally coupled, potentially self-interacting and static scalar fields all possess trivial configurations around stationary and asymptotically flat black holes. However, the linear perturbation theory of black holes have demonstrated that time-dependent bosonic fields would have superradiant instability in the background of a Kerr black hole \cite{Press:1972zz,Dolan:2007mj,Witek:2012tr,East:2017mrj} or the charged black hole \cite{Bekenstein:1973ur,Hod:2012wmy,Sanchis-Gual:2015lje}. Considering the backreaction and the conservation of the energy, the final state of a black hole with these superradiant instabilities is described by the black hole with the synchronized hair \cite{Herdeiro:2014goa,Sanchis-Gual:2015lje,East:2017ovw}. Besides, another interesting class of scalar hairy black holes were also obtained with proper conditions \cite{Herdeiro:2015waa}, such as the scalarized charged black hole in Einstein-Maxwell-Dilation gravity \cite{Gregory:1992kr,Yamazaki:1989hy,Campbell:1991kz,Garfinkle:1990qj,Kanti:1995vq,Guo:2021zed,Yao:2021zid,Zhang:2021ybj,Zhang:2021nnn,Burrage:2023zvk}, the scalarized black hole in Einstein-scalar-Gauss-Bonnet gravity \cite{Sotiriou:2013qea,Sotiriou:2014pfa,Benkel:2016kcq,Doneva:2017bvd,Silva:2017uqg,Andreou:2019ikc,Hod:2019pmb,Peng:2019snv,Ripley:2019aqj,Hod:2019vut,Dima:2020yac,Tang:2020sjs,Liu:2020yqa,Guo:2020sdu,Herdeiro:2020wei,East:2021bqk,Zhang:2022kbf}, the scalarized black hole in Einstein-Born-Infeld-scalar gravity \cite{Wang:2020ohb} and so on.

Note that, the solution of a scalarized black hole only describes the final states of the scalarization and does not show the dynamical formation of the scalar hair. To understand the dynamical formation of a scalarized black hole, one must evolve the spacetime and fields \cite{Sanchis-Gual:2015lje,Zhang:2021ybj,Zhang:2021nnn,Benkel:2016kcq,Zhang:2022kbf,Julie:2020vov,Witek:2020uzz,East:2020hgw,Doneva:2021tvn,East:2017ovw,Benkel:2016rlz,Witek:2018dmd,Doneva:2020nbb,Doneva:2021dcc,Doneva:2021dqn,East:2021bqk}. For the scalar field, the corresponding dynamical properties will be closely dependent on the self-interaction. It has been proved that the self-interaction of a scalar field also affects the properties of the scalarized black hole \cite{Gregory:1992kr,Delgado:2020hwr,Doneva:2019vuh,Macedo:2019sem,Herdeiro:2015tia,Doneva:2020kfv,Zhang:2022kbf}.

The axionic particle is one kind of candidate of dark matter and it has attracted lots of attentions \cite{Marsh:2015xka}. As a potential candidate for dark matter, the gravitational properties of the axionic particle are crucial for understanding its nature. Therefore, we will consider the scalarized black hole as one kind of possible model that black holes and dark matter can coexist. The axionic particle is inspired from the QCD axion in solving the CP problem \cite{Peccei:1977hh,tHooft:1976snw,Jackiw:1976pf}, it possesses a self-interaction potential described by the mass parameter $m$ and decay constant $f_a$ \cite{Weinberg:1977ma,Wilczek:1977pj,GrillidiCortona:2015jxo,Delgado:2020udb,Delgado:2020hwr}.

Focusing on the properties of the axionic potential, the goal of the present paper is to study gravitational effects of such a potential on the formation of a charged black hole with static scalar hair. Our paper is organized as follows. In Sec.~\ref{sec:fundamentals}, we briefly review the Einstein-Maxwell-dilaton gravity and derive the corresponding equations for the gravitational fields and matter fields. In Sec.~\ref{sec:results}, we provide the numerical results and the corresponding analysis. Finally, we give a brief conclusion and outlook in Sec.~\ref{Conclusion}.

\section{Setup} \label{sec:fundamentals}
Firstly, we give a brief introduction of the model. The corresponding action is \cite{Garfinkle:1990qj,Zhang:2021ybj}
\begin{equation}
S = \int d^4x \sqrt{|g|}
	\bigg(\frac{R-F(\phi)\mathcal{I}}{16\pi}-\frac{1}{2}\partial_\mu\phi \partial^\mu\phi -V(\phi)\bigg),
    \label{action}
\end{equation}
where we consider $F(\phi)=e^{\eta \phi}$ and $\eta$ is the coupling parameter. The invariant $\mathcal{I}$ is
\begin{equation}
\mathcal{I}=F_{\alpha\beta}F^{\alpha\beta},
\end{equation}
where $F_{\alpha\beta}=\nabla_\alpha A_\beta-\nabla_\beta A_\alpha$. The scalar potential $V(\phi)$ is taken as the axion potential \cite{GrillidiCortona:2015jxo,Delgado:2020hwr,Delgado:2020udb}:
	\begin{equation}
	V(\phi)=\frac{m^2 f_a^2}{B}\left[1-\sqrt{1-4B\sin^2\left(\frac{\phi}{2f_a}\right)}\right].
	\label{scalar_potential}
	\end{equation}
Here $m$ and $f_a$ are the mass parameter and  coupling parameter, respectively. The parameter $B(\approx 0.22)$ is determined by the ratio between the up quark and down quark masses \cite{GrillidiCortona:2015jxo}. Here, we have set the units with $G=c=\hbar=1$.

Varying the action \eqref{action} with respect to $g_{\mu\nu}$, $A_\mu$, and $\phi$, we derive the field equations as follows
\begin{equation}
R_{\mu\nu}-\frac{1}{2}g_{\mu\nu}R = 8\pi \left(T_{\mu\nu}^{(\phi)} + T_{\mu\nu}^{(\text{em})}\right),
\label{einsteineq}
\end{equation}
\begin{equation}
\nabla_\mu\left(e^{\eta \phi}F^{\mu\nu}\right) = 0,
\label{maxwelleq}
\end{equation}
\begin{equation}
\nabla_\mu\nabla^\mu\phi = \frac{\eta}{16\pi}\mathcal{I}e^{\eta \phi} + \frac{d V(\phi)}{d\phi}.
\label{keleineq}
\end{equation}
The energy-momentum tensors of the scalar field $T^{(\phi)}_{\mu\nu}$ and the electromagnetic field $T_{\mu\nu}^{(\text{em})}$ are
\begin{equation}
T^{(\phi)}_{\mu\nu} = \partial_\mu\phi\partial_\nu\phi - g_{\mu\nu}\left(\frac{1}{2}\partial^\alpha\phi\partial_\alpha\phi + V(\phi)\right)
\end{equation}
and
\begin{equation}
T_{\mu\nu}^{(\text{em})} = \frac{1}{4\pi}e^{\eta \phi}\left(F_{\mu\alpha}F_\nu^{~\alpha}-\frac{1}{4}g_{\mu\nu}F^{\alpha\beta}F_{\alpha\beta}\right).
\end{equation}
Obviously, if $\eta=0$, the above model reduces to the Einstein-Maxwell model with a minimally coupled scalar field.

It has been proved that the condition $dF(\phi)/d\phi=0$ can allow the existence of the un-scalarized bald black hole solutions \cite{Herdeiro:2018wub,Fernandes:2019rez}. For our choice of $F(\phi)=e^{\eta \phi}$, it does not satisfy the condition $dF(\phi)/d\phi=0$ and there is no static un-scalarized black hole solution in our model. However, the un-scalarized charged black hole can be considered as an initial state when we investigate the dynamical scalarization.

In this work, we mainly focus on the dynamical scalarization of a charged black hole, for which we need to carry out the nonlinear simulation of spacetime and fields. To realize the nonlinear simulation of the system described by Eqs. \eqref{einsteineq}, \eqref{maxwelleq}, and \eqref{keleineq}, we use our spherical numerical relativity code \cite{Zhang:2023qag} in terms of the Baumgarte–Shapiro–Shibata–Nakamura formalism in spherical coordinates~\cite{Montero:2012yr}. We consider the following metric
\begin{eqnarray}
ds^2&=&(-\alpha^2 + \beta^r \beta_r) dt^2 + 2\beta_r dt dr \nonumber\\
&&+e^{4\chi}\left(a\,dr^2+b\,r^2 \,d\Omega^2 \right)
\label{metric}
\end{eqnarray}
with
\begin{equation}
\beta^i=(\beta^r, 0, 0)
\end{equation}
and
\begin{equation}
d\Omega^2=d\theta^2+\sin^2\theta d\varphi^2,
\end{equation}
where $\alpha$, $\beta^i$, and $\chi$ are the lapse function, shift vector,  and conformal factor, respectively. The other functions $a$ and $b$ are the metric functions. Because we only consider the spherical case, all the functions are dependent on $(r, t)$. By using the metric \eqref{metric}, we have a 3-metric of the spacelike hypersurface $\Sigma_t$ as follows
\begin{equation}
\gamma_{ij}=e^{4\chi}\text{diag}\left(a, b\,r^2, b\,r^2\sin^2\theta\right).
\end{equation}

For simplicity, we only give a brief introduction of the evolution equations for the scalar and electromagnetic fields. For the evolution of the spacetime, the details can be found in Ref. \cite{Montero:2012yr}. For the scalar field $\phi$ described by Eq. \eqref{keleineq}, we define the conjugate momentum as
\begin{equation}
\Pi=n^\mu\nabla_{\mu}\phi,
\label{cojugate_momentum_sf}
\end{equation}
where $n^\mu$ is the normal vector of the spacelike hypersurface $\Sigma_t$. Combining the conjugate momentum \eqref{cojugate_momentum_sf}, metric \eqref{metric}, and equation of motion of the scalar field \eqref{keleineq}, we get two first-order equations \cite{Torres:2014fga,Corelli:2021ikv}:
\begin{equation}
\partial_t\phi=\beta^r\partial_r\phi + \alpha \Pi,
\end{equation}
\begin{eqnarray}
\partial_t\Pi &=& \beta^r\partial_r\Pi + \alpha \Pi K + \frac{\partial_r \phi \partial_r\alpha}{a e^{4\chi}}+ \frac{1}{2}\eta\alpha a e^{4\chi} \left(E^r\right)^2e^{\eta\phi} \nonumber\\
              && + \frac{\alpha}{a e^{4\chi}}\partial_r\phi\left(\frac{2}{r}-\frac{\partial_r a}{2a} + \frac{\partial_r b}{b} + 2 \partial_r\chi\right)\nonumber\\
              && + \frac{\alpha}{a e^{4\chi}}\partial^2_r\phi - \frac{dV(\phi)}{d\phi}.
\end{eqnarray}
Here $K$ is the trace of the extrinsic curvature $K_{ij}$.

For the electromagnetic field, the corresponding evolution equations are formulated by the electric and magnetic fields measured by the Eulerian observers \cite{Alcubierre:2009ij}:
\begin{equation}
E^\mu=-n_\mu F^{\mu\nu},
\end{equation}
\begin{equation}
B^\mu=-n_\mu {F^{*}}^{\mu\nu}.
\end{equation}
Here, ${F^{*}}^{\mu\nu}=\frac{1}{2}\epsilon^{\mu\nu\alpha\beta}F_{\alpha\beta}$. We take the convention that $\epsilon^{0123}=-1/\sqrt{-g}$ and $\epsilon_{0123}=+\sqrt{-g}$, where $g$ is the determinant of the four-dimensional metric.

Due to the spherical symmetry, the magnetic field vanishes, for which the only nonvanishing components of the electric field and vector electromagnetic potential are the radial parts $E^r$ and $a^r$. Here, $a^r$ is the radial component of the vector  electromagnetic potential $a^i=~^{(3)}A^i=\gamma^i_\mu A^\mu$. Besides, we also need to evolve the scalar potential of the electromagnetic potential $\Phi=-n_{\mu}A^\mu$. Combining with the Lorentz gauge $\nabla_\mu A^\mu=0$, we get the evolution equations of the electromagnetic field \cite{Torres:2014fga,Corelli:2021ikv}:
\begin{equation}
\partial_t E^r = \beta^r\partial_r E^r + \alpha K E^r- E^r\partial_r \beta^r - \eta\alpha \Pi E^r,
\end{equation}
\begin{equation}
\partial_t a_r = \beta^r\partial_r a_r - a_r\partial_r \beta^r - \partial_r(\alpha\varphi)-\alpha a e^{4\chi} E^r,
\end{equation}
\begin{eqnarray}
\partial_t \Phi &=&- \frac{\alpha}{a e^{4\chi}}\Bigg[\partial_r a_r + a_r\left(\frac{2}{r} - a_r \frac{\partial_r a}{2a} + \frac{\partial_r b}{b} + 2\partial_r \chi\right)\Bigg]  \nonumber\\
&&+\beta^r\partial_r\Phi + \alpha\Phi K -\frac{(\partial_r \alpha) a_r}{a e^{4\chi}}.
\end{eqnarray}
The lapse function $\alpha$ is evolved with the $1+\text{log}$ condition \cite{Bona:1997hp}
\begin{equation*}
\partial_t \alpha = - 2 \alpha K,
\end{equation*}
and the shift vector $\beta^r$ is evolved with a variation of the gamma-driver condition \cite{Alcubierre:2002kk}
\begin{equation}
\partial_t\beta^r = B^r, ~~\partial_t B^r = \frac{3}{4}\partial_t \hat{\Delta}^r - 2B^r.
\end{equation}

The scalar field vanishes at the initial time and the corresponding spacetime is completely described by a charged black hole with mass $M_0$ and charge $Q$. We take the mass and charge of the black hole at the initial time as $M_0=1$ and $Q=0.9$. The conformal electric potential $\varphi$ and conformal factor $e^{\chi}$ are \cite{Alcubierre:2009ij}
\begin{equation}
\varphi=\frac{Q}{r},
\end{equation}
\begin{equation}
e^{2\chi}=\left[\left(1+\frac{M_0}{2r}\right)^2-\frac{Q^2}{4^2}\right].
\end{equation}
The corresponding physical electric field is
\begin{equation}
E^r=\frac{Q}{r^2 e^{6\chi}}.
\end{equation}
Under the above setup, the metric functions $a=b=1$.

Combining with the units $G=c=\hbar=1$, we define the characteristic length $l_c=G M_0/c^2=M_0$, the Compton wavelength $\lambda_c=\hbar/m c = 1/m$, for which we define the dimensionless mass parameter $\bar{m}=l_c/\lambda_c = m M_0$. The dimension of the parameter $f_a$ is the same as the scalar field $\phi$, and $f_a\eta$ is dimensionless. To understand how the scalar potential \eqref{scalar_potential} affects the dynamical properties of the scalarization of the charged black hole, we consider the setup for the dimensionless parameters $\eta$, $m$, and $f_a$ as follows
\begin{eqnarray}
\bar{\eta}&=&\eta \phi_0=\left(5, 10, 15, 20\right),\\
\bar{m}&=&m M_0=\left(0, 0.25, 0.50, 0.75, 1.00\right),\\
\bar{f_a}&=&f_a/\phi_0=\left(0.03, 0.05, 0.10, 0.50, 1.00\right).
\end{eqnarray}
Here we define a constant $\phi_0$ with the dimension of the scalar field and set the value to be one.
\section{Results} \label{sec:results}

To extract the information of the system, we monitor the information of the apparent horizon and the value of the scalar hair on the apparent horizon. For a scalarized charged black hole, its properties are completely described by the irreducible mass $M_{\text{ah}}$, the value of the scalar field on the apparent horizon $\phi_{\text{ah}}$, and the charge $Q$. We have set the value of charge $Q=0.9$, therefore, tracking the values of $M_{\text{ah}}$ and $\phi_{\text{ah}}$ can completely describe the solution of the scalarized charged black hole.

Before providing the results, we go back to the scalar potential \eqref{scalar_potential}:
\begin{equation}
V(\phi)=\frac{m^2 f_a^2}{B}\left[1-\sqrt{1-4B\sin^2\left(\frac{\phi}{2f_a}\right)}\right].
\label{potential_2}
\end{equation}
Expanding the scalar potential \eqref{potential_2} around $\phi=0$, we have \cite{Delgado:2020hwr}
\begin{eqnarray}
V(\phi) &=& m^2\phi^2 - \left(\frac{3B-1}{12}\right)\frac{m^2}{f_a^2}\phi^4\nonumber\\
&&+ \frac{1+15B(3B-1)}{360f_a^4}m^2\phi^6+\cdots.
\label{expansion of scalar potential}
\end{eqnarray}
It is easy to see that $m$ and $f_a$ are the mass parameter and the quartic self-interaction coupling parameter.

\subsection{The case of $f_a\to \infty$}

\begin{figure*}[htbp]
	\begin{center}
		\includegraphics[width=\linewidth]{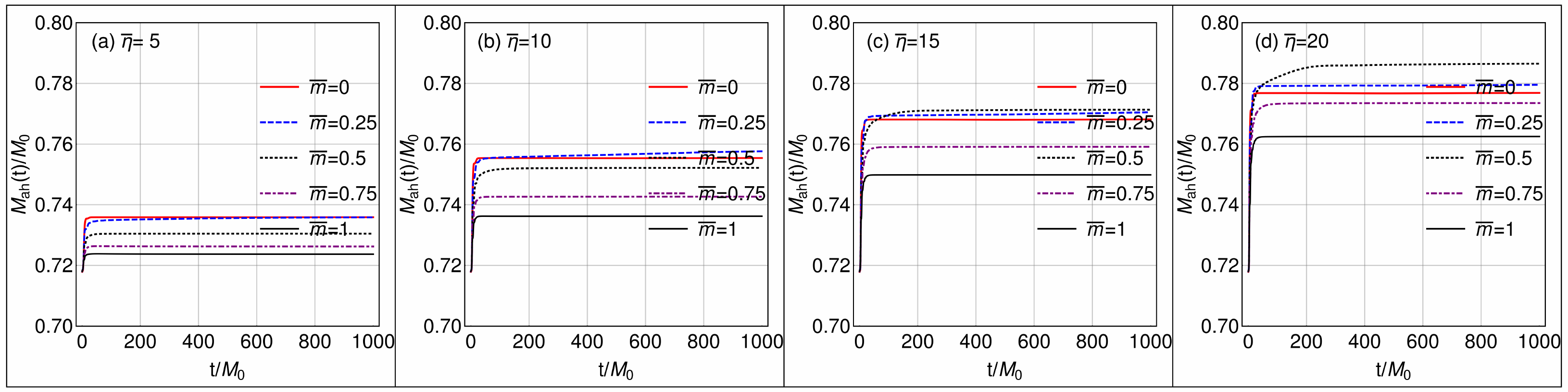}
	\end{center}
	\caption{Evolution of the irreducible mass $M_{\text{ah}}$ of the black hole with different values of the mass parameter $m$ and the coupling parameter $\eta$ for $f_a\to\infty$.}
	\label{p_m_ah}
\end{figure*}

\begin{figure*}[htbp]
	\begin{center}
		\includegraphics[width=\linewidth]{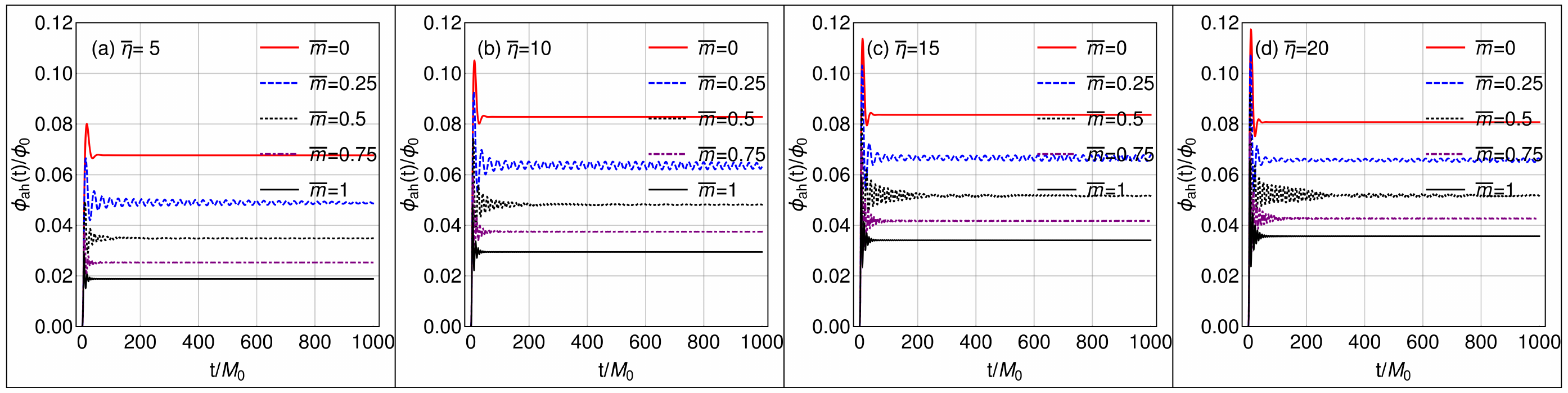}\hspace{5mm}
	\end{center}
	\caption{Evolution of the scalar field  on the apparent horizon $\phi_{\text{ah}}$ with different values of the mass parameter $m$ and the coupling parameter $\eta$ for $f_a\to\infty$.}
	\label{p_phi_ah}
\end{figure*}

Note that the scalar potential \eqref{expansion of scalar potential} degenerates into the massive scalar field case with a sufficient large $f_a$ ($f_a\gg \phi$). Firstly, we focus on the case that $f_a\to\infty$ and study how the mass term of the scalar field affects the properties of the scalarization. Figures \ref{p_m_ah} and \ref{p_phi_ah} show the irreducible mass $M_{\text{ah}}$ and the value of the scalar field $\phi_{\text{ah}}$ on the apparent horizon as functions of time. We can see that both the $M_{\text{ah}}$ and $\phi_{\text{ah}}$ will relax to the fixed values described by the stationary scalarized solutions. Such behaviors are similar to the cases in Refs. \cite{Zhang:2021ybj,Zhang:2022kbf,Benkel:2016kcq}.

For the irreducible masses of the black holes given in Fig. \ref{p_m_ah}, we observe that all of them increase with time and finally relax to fixed values when the unscalarized charged black hole transforms to the scalarized black hole. For the values of the scalar field on the apparent horizon $\phi_{\text{ah}}$ given in Fig. \ref{p_phi_ah}, we find that they increase rapidly in the early stage, and after reaching the maximal values, they start to decrease and finally relax to the fixed values. We also observe that when the mass parameter $m$ is zero or with large values, $\phi_{\text{ah}}$ quickly relaxes to a fixed value. For the case with an intermediate mass parameter $m$, the values of the scalar field on the apparent horizon oscillate when they tend to the final state. Such behaviors are similar to the case of \cite{Zhang:2022kbf,Benkel:2016kcq}.

\begin{figure}[htbp]
	\begin{center}
		\includegraphics[width=\linewidth]{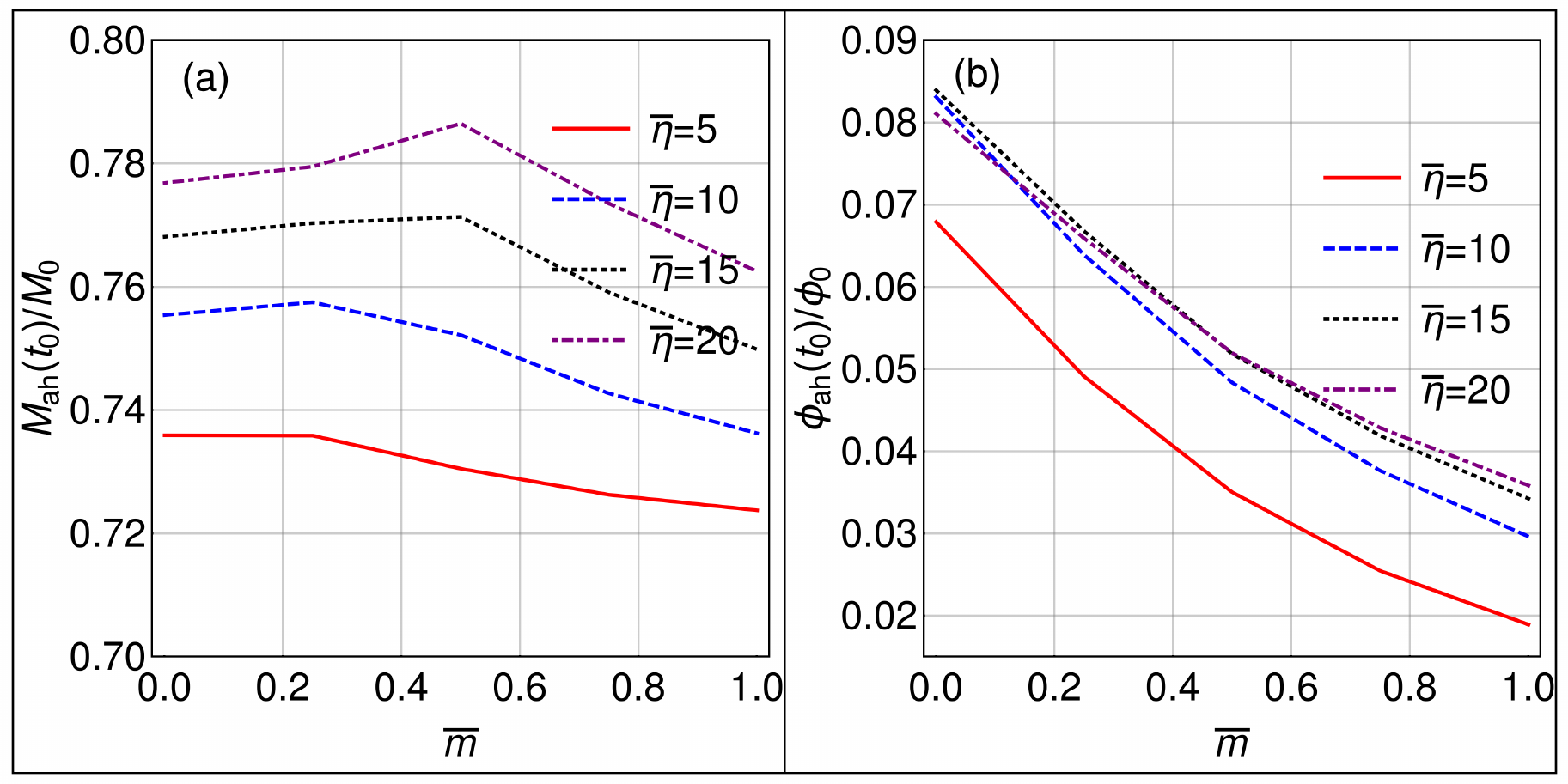}
	\end{center}
	
	\caption{The irreducible mass $M_{\text{ah}}$ of the black hole and the values of dilaton field $\phi_{\text{ah}}$ on the apparent horizon with different values of the mass parameter $m$ and the coupling parameter $\eta$ at late time $t_0=900M$.}
	\label{m_phi_ah_latetime_finfty}
\end{figure}

The simulation results indicate that both the values of the irreducible mass $M_{\text{ah}}$ and scalar field $\phi_{\text{ah}}$ will relax to fixed values at late time that describe the static hairy black hole. Therefore, to show how the mass parameter $m$ and coupling parameter $\eta$ affect the final state of the scalarized black hole, we extract the values of $M_{\text{ah}}$ and $\phi_{\text{ah}}$ at late time $t_0=900 M_0$. Due to the oscillation behavior of the scalar field $\phi_{\text{ah}}$, we extract the averaged values at late time. We plot the values extracted at the late time in Fig. \ref{m_phi_ah_latetime_finfty}.

Comparing the results with different values of the mass parameter $m$ and coupling parameter $\eta$, we can see that when the coupling parameter $\eta$ is small, the irreducible mass of the final hairy black hole decreases with the mass parameter $m$. While the coupling parameter $\eta$ increases, the irreducible mass of the final scalarized black hole will not monotonically decrease with the mass parameter $m$, see the details in Fig. \ref{m_phi_ah_latetime_finfty}. But for the value of the scalar field $\phi_{\text{ah}}$ on the apparent horizon, we find that when the coupling parameter $\eta$ changes, the value of the scalar field on the apparent horizon always decreases with the mass parameter $m$.
	
\subsection{The case of finite $f_a$}

Next, we move to the cases where $f_a$ takes finite values. We have checked that in the process of the dynamical scalarization of the charged black hole, the changes of the irreducible mass $M_{\text{ah}}$ and the value of the scalar field $\phi_{\text{ah}}$ on the apparent horizon have the similar dynamical behaviors as shown in Figs. \ref{p_m_ah} and \ref{p_phi_ah}. Therefore, we only provide the values of $M_{\text{ah}}$ and $\phi_{\text{ah}}$ at late time $t_0= 900 M_0$.

\begin{figure*}[htbp]
	\begin{center}
		\includegraphics[width=\linewidth]{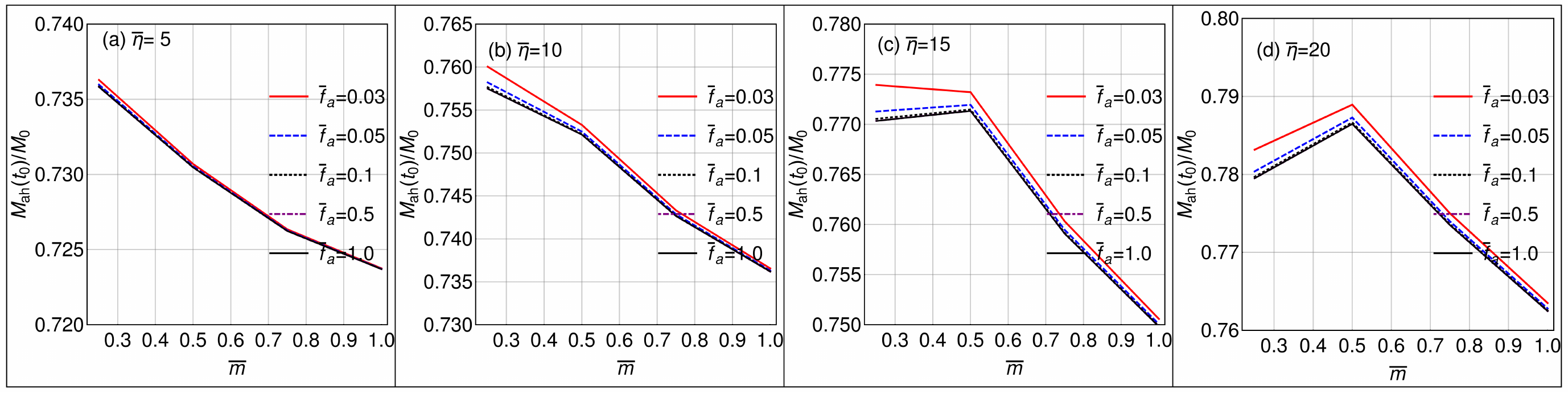}
	\end{center}
	\caption{The irreducible mass $M_{\text{ah}}/{M_0}$ of the black hole with different values of the parameters $m$, $f_a$, and $\eta$.}
	\label{p_m_ah_varmfa}
\end{figure*}

Figure \ref{p_m_ah_varmfa} shows the irreducible mass $M_{\text{ah}}$ of the black hole with different values of parameters $m$, $f_a$, and $\eta$. Here, we observe the similar relations between the irreducible mass $M_{\text{ah}}$ and the mass parameter $m$ of the scalar field with fixed $f_a$ and $\eta$. It can be seen that, for the case with a fixed parameter $f_a$ and a small coupling parameter $\eta$, the irreducible mass of the black hole decreases with the mass parameter $m$. The irreducible mass of the black hole will not monotonically decrease with the mass parameter $m$ when the coupling parameter $\eta$ increases. To see how the irreducible mass $M_{\text{ah}}$ depends on the parameter $f_a$, we check the changes of $M_{\text{ah}}$ with different values of $f_a$ with fixed values of $m$ and $\eta$. Noticing that the tiny differences of the irreducible mass $M_{\text{ah}}$ in Fig. \ref{p_m_ah_varmfa} can not clearly show the details with a large mass parameter $m$, we list the corresponding values of the irreducible mass $M_{\text{ah}}$ in Tables \ref{values_of_mah_a}, \ref{values_of_mah_b}, \ref{values_of_mah_c}, and \ref{values_of_mah_d}. We find that the irreducible mass $M_{\text{ah}}$ of the black hole decreases slightly with the parameter $f_a$.

\begin{table}[!htb]
	\begin{center}
		\caption{Values of the irreducible mass $M_{\text{ah}}/{M_0}$ of the black hole with different values of the parameters $\bar{m}$ and $\bar{f}_a$ and $\bar{\eta}=5$ at late time $t_0 = 900 M_0$.}
		\begin{tabular}{ c| c  c  c  c  c}
			\diagbox{$\bar{m}$}{$\bar{f}_a$} &~~0.03~   ~&~~0.05~   ~&~~0.10~~&~~0.50~~&~~1.00~~\\
			\hline
			~~0.25~              &~0.73629~&~0.73598~&~0.73586~&~0.73582~&~0.73582~\\
			\hline	
			~~0.50~              &~0.73070~&~0.73055~ &~0.73049~&~0.73047~&~0.73047~\\
			\hline
			~~0.75~              &~0.72635~&~0.72628~&~0.72625~&~0.72624~&~0.72624~\\
			\hline
			~~1.00~              &~0.72376~&~0.72372~&~0.72371~&~0.72370~&~0.72370~\\

		\end{tabular}
		\label{values_of_mah_a}
	\end{center}
\end{table}

\begin{table}[!htb]
	\begin{center}
		\caption{Values of the irreducible mass $M_{\text{ah}}/{M_0}$ of the black hole with different values of the parameters $\bar{m}$ and $\bar{f}_a$ and $\bar{\eta}=10$ at late time $t_0 = 900 M_0$.}
		\begin{tabular}{ c| c  c  c  c  c}
			\diagbox{$\bar{m}$}{$\bar{f}_a$} &~~0.03~   ~&~~0.05~   ~&~~0.10~~&~~0.50~~&~~1.00~~\\
			\hline
			~~0.25~              &~0.76005~&~0.75822~&~0.75766~&~0.75749~&~0.75749~\\
			\hline
			~~0.50~              &~0.75328~&~0.75250~ &~0.75223~&~0.75215~&~0.75214~\\
			\hline
			~~0.75~              &~0.74332~&~0.74287~&~0.74271~&~0.74266~&~0.74266~\\
			\hline
			~~1.00~              &~0.73657~&~0.73632~&~0.73623~&~0.73620~&~0.73620~\\
		\end{tabular}
		\label{values_of_mah_b}
	\end{center}
\end{table}

\begin{table}[!htb]
	\begin{center}
		\caption{Values of the irreducible mass $M_{\text{ah}}/{M_0}$ of the black hole with different values of the parameters $\bar{m}$ and $\bar{f}_a$ and $\bar{\eta}=15$ at late time $t_0 = 900 M_0$.}
		\begin{tabular}{ c| c  c  c  c  c}

			\diagbox{$\bar{m}$}{$\bar{f}_a$} &~~0.03~   ~&~~0.05~   ~&~~0.10~~&~~0.50~~&~~1.00~~\\
			\hline
			~~0.25~              &~0.77394~&~0.77129~&~0.77056~&~0.77035~&~0.77034~\\
			\hline
			~~0.50~              &~0.77322~&~0.77194~ &~0.77148~&~0.77135~&~0.77134~\\
			\hline
			~~0.75~              &~0.76033~&~0.75947~&~0.75918~&~0.75909~&~0.75909~\\
			\hline
			~~1.00~              &~0.75059~&~0.75007~&~0.74988~&~0.74982~&~0.74982~\\
		\end{tabular}
		\label{values_of_mah_c}
	\end{center}
\end{table}

\begin{table}[!htb]
	\begin{center}
		\caption{Values of the irreducible mass $M_{\text{ah}}/{M_0}$ of the black hole with different values of the parameters $\bar{m}$ and $\bar{f}_a$ and $\bar{\eta}=20$ at late time $t_0 = 900 M_0$.}
		\begin{tabular}{ c| c  c  c  c  c}
			\diagbox{$\bar{m}$}{$\bar{f}_a$} &~~0.03~   ~&~~0.05~   ~&~~0.10~~&~~0.50~~&~~1.00~~\\
			\hline
			~~0.25~              &~0.78318~&~0.78040~&~0.77971~&~0.77952~&~0.77951~\\
			\hline
			~~0.50~              &~0.78896~&~0.78730~ &~0.78668~&~0.78649~&~0.78649~\\
			\hline
			~~0.75~              &~0.77501~&~0.77398~&~0.77362~&~0.77351~&~0.77351~\\
			\hline
			~~1.00~              &~0.76352~&~0.76281~&~0.76256~&~0.76248~&~0.76248~\\
		\end{tabular}
		\label{values_of_mah_d}
	\end{center}
\end{table}

\begin{figure*}[htbp]
	\begin{center}
		\includegraphics[width=\linewidth]{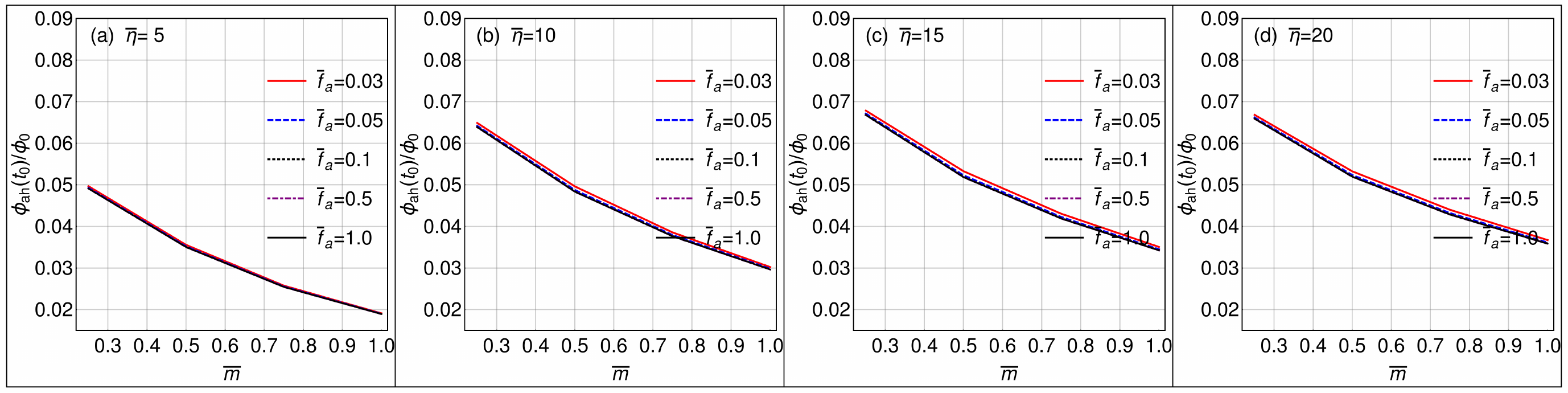}\hspace{5mm}
	\end{center}
	\caption{Values of the scalar field $\phi_{\text{ah}}$ on the apparent horizon with different values of the parameters $m$, $f_a$, and $\eta$.}
	\label{p_phi_ah_varmfa}
\end{figure*}

We also provide the results about how the value of the scalar field on the apparent horizon $\phi_{\text{ah}}$ is affected by the coupling parameter $f_a$ in the axionic potential, see the details in Fig. \ref{p_phi_ah_varmfa}. Due to the tiny difference of the value of the scalar field on the apparent horizon $\phi_{\text{ah}}$ in Fig. \ref{p_phi_ah_varmfa}, we also list the concrete values of $\phi_{\text{ah}}$ in Tables \ref{values_of_phiah_a}, \ref{values_of_phiah_b}, \ref{values_of_phiah_c}, and \ref{values_of_phiah_d}. We find that $\phi_{\text{ah}}$ monotonically decreases slightly with the coupling parameter $f_a$ with fixed values of $m$ and $\eta$.

\begin{table}[!htb]
	\begin{center}
		\caption{Values of the scalar field on the apparent horizon $\phi_{\text{ah}}/\phi_0$ with different values of the parameters $\bar{m}$ and $\bar{f}_a$ and $\bar{\eta}=5$ at late time $t_0 = 900 M_0$.}
		\begin{tabular}{ c| c  c  c  c  c}

			\diagbox{$\bar{m}$}{$\bar{f}_a$} &~~0.03~   ~&~~0.05~   ~&~~0.10~~&~~0.50~~&~~1.00~~\\
			\hline
			~~0.25~              &~0.04969~&~0.04931~&~0.04916~&~0.04913~&~0.04911~\\
			\hline
			~~0.50~              &~0.03557~&~0.03522~ &~0.03508~&~0.03503~&~0.03503~\\
			\hline
			~~0.75~              &~0.02578~&~0.02557~&~0.02549~&~0.02546~&~0.02546~\\
			\hline
			~~1.00~              &~0.01911~&~0.01900~&~0.01895~&~0.01894~&~0.01894~\\

		\end{tabular}
		\label{values_of_phiah_a}
	\end{center}
\end{table}

\begin{table}[!htb]
	\begin{center}
		\caption{Values of the scalar field on the apparent horizon $\phi_{\text{ah}}/\phi_0$ with different values of the parameters $\bar{m}$ and $\bar{f}_a$ and $\bar{\eta}=10$ at late time $t_0 = 900 M_0$.}
		\begin{tabular}{ c| c  c  c  c  c}

			\diagbox{$\bar{m}$}{$\bar{f}_a$} &~~0.03~   ~&~~0.05~   ~&~~0.10~~&~~0.50~~&~~1.00~~\\
			\hline
			~~0.25~              &~0.06491~&~0.06421~&~0.06397~&~0.06389~&~0.06389~\\
			\hline
			~~0.50~              &~0.04964~&~0.04878~ &~0.04846~&~0.04836~&~0.04836~\\
			\hline
			~~0.75~              &~0.03857~&~0.03796~&~0.03772~&~0.03765~&~0.03765~\\
			\hline
			~~1.00~              &~0.03024~&~0.02985~&~0.02970~&~0.02965~&~0.02965~\\

		\end{tabular}
		\label{values_of_phiah_b}
	\end{center}
\end{table}

\begin{table}[!htb]
	\begin{center}
		\caption{Values of the scalar field on the apparent horizon $\phi_{\text{ah}}/\phi_0$ with different values of the parameters $\bar{m}$ and $\bar{f}_a$ and $\bar{\eta}=15$ at late time $t_0 = 900 M_0$.}
		\begin{tabular}{ c| c  c  c  c  c}

			\diagbox{$\bar{m}$}{$\bar{f}_a$} &~~0.03~   ~&~~0.05~   ~&~~0.10~~&~~0.50~~&~~1.00~~\\
			\hline
			~~0.25~              &~0.06786~&~0.06714~&~0.06687~&~0.06679~&~0.06679~\\
			\hline
			~~0.50~              &~0.05329~&~0.05231~ &~0.05196~&~0.05185~&~0.05185~\\
			\hline
			~~0.75~              &~0.04307~&~0.04229~&~0.04199~&~0.04190~&~0.04190~\\
			\hline
			~~1.00~              &~0.03508~&~0.03453~&~0.03432~&~0.03425~&~0.03425~\\

		\end{tabular}
		\label{values_of_phiah_c}
	\end{center}
\end{table}

\begin{table}[!htb]
	\begin{center}
		\caption{Values of the scalar field on the apparent horizon $\phi_{\text{ah}}/\phi_0$ with different values of the parameters $\bar{m}$ and $\bar{f}_a$ and $\bar{\eta}=20$ at late time $t_0 = 900 M_0$.}
		\begin{tabular}{ c| c  c  c  c  c}

			\diagbox{$\bar{m}$}{$\bar{f}_a$} &~~0.03~   ~&~~0.05~   ~&~~0.10~~&~~0.50~~&~~1.00~~\\
			\hline
			~~0.25~              &~0.06685~&~0.06623~&~0.06599~&~0.06592~&~0.06591~\\
			\hline
			~~0.50~              &~0.05326~&~0.05237~ &~0.05206~&~0.05196~&~0.05196~\\
			\hline
			~~0.75~              &~0.04402~&~0.04324~&~0.04295~&~0.04286~&~0.04286~\\
			\hline
			~~1.00~              &~0.03673~&~0.03615~&~0.03593~&~0.03586~&~0.03585~\\

		\end{tabular}
		\label{values_of_phiah_d}
	\end{center}
\end{table}

\section{Conclusions}\label{Conclusion}

We reported on the dynamical formation of the axionic hair surrounding a charged black hole. This investigation was conducted by postulating an exponential coupling between the scalar field and the electromagnetic field, as suggested by low-energy predictions from string theory \cite{Garfinkle:1990qj}. Our focus was on understanding how a scalar hair could dynamically form, driven by the invariant term $\mathcal{I}=F_{\alpha\beta}F^{\alpha\beta}$ within the context of a charged black hole's background.

For a scalarized black hole, its properties are fully encapsulated by the irreducible mass $M_{\text{ah}}$, the value of the scalar field on the apparent horizon $\phi_{\text{ah}}$, and the charge $Q$, given fixed parameter values for $m$ and $f_a$. We monitored the behavior of the scalar field at various stages and observed that all configurations of the scalar field converge to static states. This outcome aligns with findings reported in Refs. \cite{Zhang:2021ybj,Zhang:2022kbf,Benkel:2016kcq}. Furthermore, we elucidated the relationships between these variables and the axionic potential under different values of $m$ and $f_a$, to demonstrate how the characteristics of the scalarized black hole are influenced by the axionic potential.

When the decay constant $f_a$ is infinite, the axionic potential degenerates into a mass term. We explored how this mass term influences the dynamical scalarization of a charged black hole. Our findings indicate that the irreducible mass $M_{\text{ah}}$ of the resulting scalarized black hole decreases monotonically with the mass parameter when the coupling constant $\eta$ is small. However, for larger values of the coupling constant $\eta$, the irreducible mass $M_{\text{ah}}$ does not consistently decrease with the mass parameter. In contrast, the value of the scalar field on the apparent horizon $\phi_{\text{ah}}$ does not consistently diminishes with the mass parameter as the coupling constant transitions from small to large values.

Furthermore, we examined the effects of a finite decay constant $f_a$ on the dynamical scalarization of a charged black hole. Our results showed that both the irreducible mass $M_{\text{ah}}$ and the scalar field on the apparent horizon $\phi_{\text{ah}}$ of the final black hole decrease slightly with an increasing decay constant $f_a$.

We presented these findings to provide insights into how the axionic potential affects the dynamical scalarization of charged black holes, contributing to a deeper understanding of this phenomenon.
	
\acknowledgments

This work was supported in part by the National Key Research and Development Program of China (Grant No. 2021YFC2203003), the National Natural Science Foundation of China (Grants No. 12105126, No. 12475056, No. 12305065, No. 12247178, No. 12075103, and No 12247101), the 111 Project under (Grant No. B20063), the China Postdoctoral Science Foundation (No. 2023M731468), the Gansu Province's Top Leading Talent Support Plan.

\end{document}